\documentclass[conference]{IEEEtran}
\IEEEoverridecommandlockouts

\usepackage{ifthen}
\newboolean{isanonymous}
\setboolean{isanonymous}{false} 

\newboolean{useieeebottomfooter}
\setboolean{useieeebottomfooter}{false} 

\makeatletter
\@for\opt:=\@classoptionslist\do{%
    \ifthenelse{\equal{\opt}{anonymous}}{\setboolean{isanonymous}{true}}{}
}
\makeatother

\newcommand{\blind}[1]{%
    \ifbool{isanonymous}{\textit{Blinded for review}}{#1}%
}

\usepackage[hidelinks]{hyperref}
\newcommand{\Autoref}[1]{%
  \begingroup%
  \renewcommand{\sectionautorefname}{Section}%
  \renewcommand{\subsectionautorefname}{Subsection}%
  \renewcommand{\subsubsectionautorefname}{Subsubsection}%
  \renewcommand{\figureautorefname}{Figure}%
  \renewcommand{\tableautorefname}{Table}%
  \autoref{#1}%
  \endgroup%
}

\usepackage{tikz}
\usetikzlibrary{shapes.geometric, arrows}

\usepackage[numbers]{natbib}
\usepackage{amsmath,amssymb,amsfonts}
\usepackage{algorithmic}
\usepackage{graphicx}
\usepackage{textcomp}

\usepackage{tabularx}
\usepackage{multirow}
\usepackage{multicol}

\usepackage{helvet} 
\usepackage{fancyhdr}

\usepackage{url} 

\usepackage[acronym]{glossaries}

\newacronym{CCPA}{CCPA}{California Consumer Privacy Act}

\newacronym{AI}{AI}{Artificial Intelligence}

\newacronym{DP}{DP}{Differential Privacy}

\newacronym{SecAgg}{SecAgg}{Secure Aggregation}

\newacronym{SMPC}{SMPC}{Secure Multiparty Computation}

\newacronym{IZ}{IT}{Information Technology}

\newacronym{FI}{FI}{Financial Institution}

\newacronym{FedAvg}{Fed-Avg}{Federated Average}

\newacronym{Fed-SGD}{Fed-SGD}{Federated Stochastic gradient descent}

\newacronym{NN}{NN}{Neural Network}

\newacronym{GPU}{GPU}{Graphics Processing Unit}

\newacronym{SME}{SME}{Small and Medium-sized Enterprise}

\newacronym{CSR}{CSR}{Case Study Research}

\newacronym{GT}{GT}{Grounded Theory}

\newacronym{TOE}{TOE}{Technology–Organization–Environment}

\newacronym{EU}{EU}{European Union}

\newacronym{CEP}{CEP}{Clean Energy Package}

\newacronym[plural=ICTs (technologies), longplural=Information and Communication Technologies]{ICT}{ICT}{Information and Communication Technology}

\newacronym{IIoT}{IIoT}{Industrial Internet Of Things}

\newacronym{IoT}{IoT}{Internet Of Things}

\newacronym{DLT}{DLT}{Distributed Ledger Technology}

\newacronym{SGAM}{SGAM}{Smart Grid Architecture Model}

\newacronym{DSO}{DSO}{Distribution System Operator}

\newacronym{TSO}{TSO}{Transmission System Operator}

\newacronym{LFM}{LFM}{Local Flexibility Market}

\newacronym{DER}{DER}{Distributed Energy Resource}

\newacronym{RES}{RES}{Renewable Energy Resources}

\newacronym{EV}{EV}{Electric Vehicle}

\newacronym{DR}{DR}{Demand Response}

\newacronym{HV}{HV}{High Voltage}

\newacronym{MV}{MV}{Medium Voltage}

\newacronym{LV}{LV}{Low Voltage}

\newacronym{SRA}{SRA}{Scalability and Replicability Analysis}

\newacronym{OLTC}{OLTC}{On Load Tap Changer}

\newacronym{DSM}{DSM}{Demand Side Management}

\newacronym{p2p}{p2p}{peer-to-peer}

\newacronym{ETDP}{ETDP}{extended taxonomy design process}

\newacronym{ACER}{ACER}{Agency for the Cooperation of Energy Regulators}

\newacronym{SO}{SO}{system operator}

\newacronym{LEM}{LEM}{local electricity market}

\newacronym{USEF}{USEF}{universal smart energy framework}

\newacronym{FSP}{FSP}{flexibility service provider}

\newacronym{OCPP}{OCPP}{Open Charge Point Protocol}

\newacronym{EFDM}{EFDM}{energy flexibility data model}

\newacronym{CIM}{CIM}{common information model}

\newacronym{GDPR}{GDPR}{General Data Protection Regulation}

\newacronym{NIST}{NIST}{National Institute of Standards and Technology}

\newacronym{OSI}{OSI}{Open Systems Interconnection}

\newacronym{TCP}{TCP}{Transmission Control Protocol/Internet Protocol}

\newacronym{IP}{IP}{Internet Protocol}

\newacronym{UDP}{UDP}{User Datagram Protocol}

\newacronym{IED}{IED}{Intelligent electrical device}

\newacronym{UC}{UC}{use case}

\newacronym{BUC}{BUC}{business use case}

\newacronym{MS}{MS}{Member State}

\newacronym{NECP}{NECP}{National Energy and Climate Plan}

\newacronym{ENTSOE}{ENTSO-E}{European Network of Transmission System Operators for Electricity}

\newacronym{UMEI}{UMEI}{Universal Market Enabling Interface}
 
\newacronym{MO}{MO}{market operator}

\newacronym{CIA}{CIA}{Confidentiality, Integrity and Availability}

\newacronym{OMIE}{OMIE}{Iberian Electricity Market Operator}

\newacronym{ENW}{ENWL}{Electricity North West Ltd.}

\newacronym{DPS}{DPS}{Dynamic Procurement System}

\newacronym{ES}{ES}{Spain}

\newacronym{UK}{UK}{United Kingdom}

\newacronym{NL}{NL}{Netherlands}

\newacronym{ITT}{ITT}{Invitation to Tender}

\newacronym{ENA}{ENA}{Energy Networks Association}

\newacronym{NG}{NGED}{National Grid Electricity Distribution}

\newacronym{DNO}{DNO}{Distribution Network Operator}

\newacronym{API}{API}{Application Programme Interface}

\newacronym{ETPA}{ETPA}{Energy Trading Platform Amsterdam}

\newacronym{GMS}{GMS}{Grid and Management Service}

\newacronym{GOPACS}{GOPACS}{Grid Operators Platform for Congestion Spreads}

\newacronym{JSON}{JSON}{JavaScript Object Notation}

\newacronym{IEA}{IEA}{International Energy Agency}

\newacronym{DSR}{DSR}{Design Science Research}

\newacronym{DSRM}{DSRM}{Design Science Research Methodology Process Model}

\newacronym{IS}{IS}{Information Systems}

\newacronym{RP}{RP}{Research Publication}

\newacronym{PV}{PV}{Photovoltaic}

\newacronym{SoS}{SoS}{System of Systems}

\newacronym{FL}{FL}{Federated Learning}

\newacronym{DS}{DS}{Design Science}

\newacronym{SOA}{SOA}{Service-oriented Architecture}

\newacronym{XaaS}{XaaS}{X as a Service}

\newacronym{BRP}{BRP}{Balance Responsible Party}

\newacronym{P2P}{P2P}{Peer-to-Peer}

\newacronym{SaaS}{SaaS}{Software as a Service}

\newacronym{EES}{EES}{Electrochemical Energy Storage}

\newacronym{TLS}{TLS}{Traffic Light System}

\newacronym{UML}{UML}{Unified Modeling Language}

\newacronym{SCADA}{SCADA}{Supervisory Control And Data Acquisition}

\newacronym{ADMS}{ADMS}{Advance Distribution Management System}

\newacronym{RTU}{RTU}{Remote Terminal Unit}

\newacronym{SOTA}{SOTA}{State-Of-the-Art}

\newacronym{ESP}{ESP}{Energy Synchronization Platform}

\newacronym{CP}{CP}{Company Platform}

\newacronym{MP}{MP}{Market Platform}

\newacronym{TF}{TF}{Task Force}

\newacronym{PF}{PF}{Power Flow}

\newacronym{OPF}{OPF}{Optimal Power Flow}

\newacronym{SSU}{SSU}{Smart Storage Unit}

\newacronym{MPOPF}{MPOPF}{Multi-Period Optimal Power Flow}

\newacronym{VPP}{VPP}{Virtual Power Plant}

\newacronym{HEMS}{HEMS}{Home Energy Management System}

\newacronym{ESS}{ESS}{Energy Storage System}

\newacronym{ML}{ML}{Machine Learning}

\newacronym{DL}{DL}{Deep Learning}

\newacronym{STLF}{STLF}{Short-Term Load Forecasting}

\newacronym{LSTM}{LSTM}{Long Short-Term Memory}

\newacronym{DTW}{DTW}{Dynamic Time Warping}

\newacronym{USA}{USA}{United States of America}

\newacronym{NILM}{NILM}{Non-intrusive load monitoring}

\newacronym{FCL}{FCL}{Fully Connected Layers}

\newacronym{RMSE}{RMSE}{Root Mean Square Error}

\newacronym{NRMSE}{NRMSE}{Normalized Root Mean Square Error}

\newacronym{MAPE}{MAPE}{Mean Absolute Percentage Error}

\newacronym{MAE}{MAE}{Mean Absolute Error}

\newacronym{MSE}{MSE}{Mean Squared Error}

\newacronym{MASE}{MASE}{Mean Absolute Scaled Error}

\newacronym{DBI}{DBI}{Davies–-Bouldin index}

\newacronym{CPO}{CPO}{Charging Point Operator}

\newacronym{CS}{CS}{Charging Station}

\newacronym{EE}{EE}{Energy Efficiency}

\newacronym{NREL}{NREL}{National Renewable Energy Laboratory}

\newacronym{CNN}{CNN}{Convolutional Neural Network}

\newacronym{EVCP}{EVCP}{electric vehicle Charging Pile}

\newacronym{CPU}{CPU}{Central Processing Unit}

\newacronym{PID}{PID}{proportional-integral-derivative}

\newacronym{EEM}{EEM}{European electricity market}

\newacronym[plural=LECs, longplural=Local Energy Communities]{LEC}{LEC}{Local Energy Community}

\newacronym{LM}{LM}{local market}

\newacronym{ME}{ME}{Member State}

\newacronym[plural=CECs, longplural=Citizen Energy Communities]{CEC}{CEC}{Citizen Energy Community}

\newacronym[plural=RECs, longplural=Renewable Energy Communities]{REC}{REC}{Renewable Energy Community}

\newacronym[plural=ECs, longplural=Energy Communities]{EC}{EC}{Energy Community}

\renewcommand{\figureautorefname}{Fig.}

\usepackage{orcidlink}
\newcommand{\orcid}[1]{\href{https://orcid.org/#1}{\includegraphics[width=10pt]{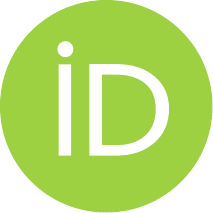}}}

\def\BibTeX{{\rm B\kern-.05em{\sc i\kern-.025em b}\kern-.08em
    T\kern-.1667em\lower.7ex\hbox{E}\kern-.125emX}}
    
\begin{document}

\title{Where do local markets fit in the current European Electricity Market structures?\thanks{
\noindent\hspace{-0.3cm}\hrulefill\\
\blind{\indent This research was funded in part by the Luxembourg National Research Fund (FNR) and PayPal, PEARL grant reference 13342933/Gilbert Fridgen, by FNR grant reference HPC BRIDGES/2022\_Phase2/17886330/DELPHI and by the Luxembourgish Ministry of Economy with grant reference 20230227RDI170010375846. Also, the authors gratefully acknowledge the financial support of Creos Luxembourg under the research project FlexBeAn. For the purpose of open access, and in fulfillment of the obligations arising from the grant agreement, the author has applied a Creative Commons Attribution 4.0 International (CC BY 4.0) license to any Author Accepted Manuscript version arising from this submission.}}
}

\ifbool{isanonymous}{
    \author{Anonymous Author(s)}
}{
\author{\IEEEauthorblockN{Sergio Potenciano Menci\textsuperscript{1}\orcid{0000-0002-9032-7183}, Laura Andolfi\textsuperscript{1}\orcid{0000-0003-1262-9511}, Rawan Akkouch\textsuperscript{1}\orcid{0009-0005-6672-3732}}
\IEEEauthorblockA{\textit{Interdisciplinary Centre for Security, Reliability and Trust - SnT}, \textit{University of Luxembourg}, Luxembourg}
}
}

\ifbool{useieeebottomfooter}{
  \fancypagestyle{firstpage}{
    \fancyhf{}
    \renewcommand{\headrulewidth}{0pt} 
    \fancyfoot[L]{\fontsize{9}{11}\fontfamily{phv}\selectfont\textbf{979-8-3315-1278-1/25/\$31.00 ©2025 IEEE}}
    \setlength{\footskip}{1.27cm}
  }
  \thispagestyle{firstpage}
}{%
  \pagestyle{plain}
}

\maketitle
\ifbool{useieeebottomfooter}{
\thispagestyle{firstpage} 
}

\begin{abstract}
\acrlongpl{EEM} have been complex since their inception. Policies and technologies advancing renewable integration, consumer empowerment, flexibility, and electrification are reshaping generation and consumption, increasing this complexity. System operators face congestion, voltage management, and redispatch challenges while market actors navigate imbalances and volatility.
New market structures, such as energy communities and local flexibility markets, aim to address local energy dynamics and integrate decentralized flexibility. However, their fit within existing market frameworks remains unclear, leading to inconsistent interpretations and regulatory uncertainties. This manuscript presents a graphical classification of local markets, positioning them within electricity procurement (e.g., wholesale) and system operation (e.g., ancillary) services while illustrating their interrelations. Despite their potential, these markets remain in early development, facing regulatory ambiguities, resource limitations, and coordination challenges.

\end{abstract}

\vspace{0.5cm}
\noindent



\sloppy

\section{Introduction}
\label{sec:intro}

Over the past two decades, the \gls{EU} has actively transformed its electricity sector through a series of energy packages. 
The primary objective remains the establishment of a \textit{"competitive, customer-oriented, flexible, and non-discriminatory"} energy market~\cite{fact_sheet_EU}.

The resulting energy market, encompassing the electricity and gas sectors, has been inherently complex since its inception, particularly within the electricity sector~\cite{8606494}. Electricity is a unique commodity that requires real-time physical balance, as electricity must be consumed as it is generated to stay in balance~\cite{collins2002economics}. Consequently, a range of markets have emerged within the \gls{EEM} to address the unique challenges of trading and managing electricity, forming a sophisticated structure that continues to evolve~\cite{7521197}. This evolution has been largely accentuated by the most recent energy packages—the fourth, \textit{Clean Energy for All Europeans}~\cite{clean_energy_package}, and the fifth, \textit{Fit for 55}~\cite{eu_fit_for_55}. These policies prioritize renewable energy integration, consumer empowerment, flexibility, and electrification across sectors such as transportation and industry~\cite{Bogdanov2020Full}.

As these policies take effect, they reshape traditional energy generation, consumption, and distribution patterns, intensifying the challenges of an already strained system~\cite{power_systems_SOTA}. For example, system operators now face increased challenges, such as managing congestion, maintaining voltage stability, and coordinating re/dispatch at more localized levels~\cite{cahllenges_power_system}. Meanwhile, market participants must engage with intensified imbalances and greater market volatility ~\cite{Heptonstall2020ASR}, often stemming from local-level disturbances caused by distributed renewable integration and electrification ~\cite{CABALLEROPENA2022107900, GAUR2022113136}.

In response to these challenges, new concepts emerge to address these new local energy dynamics and leverage decentralized flexibility through programs such as \gls{DR} ~\cite{dsm_definition}. At the same time, the \gls{EU} has consistently pushed for introducing any new solution as a market-based solution ~\cite{eu_regulation_943, european_parliament_and_of_the_council_directive_2019}. Thus, these new concepts emerge as new market structures. Notably, these are local energy communities~\cite{european_commission_joint_research_centre_energy_2020} or markets and flexibility markets~\cite{lfm_spm}.
These emerging market concepts are being explored from diverse perspectives in academic and non-academic literature (see \Autoref{sec:eem_structures}). Especially the non-academic but policy-oriented documents~\cite{eu_regulation_943, european_parliament_and_of_the_council_directive_2019} are addressing them, and upcoming ones, including the Guidelines for Demand Response~\cite{ACER_framework_guidelines} and the proposal for Network Code for Demand Response ~\cite{demand_response_network_code_proposal}. 
Nevertheless, as with any emerging topic of increasing significance and a growing body of literature, organizing knowledge presents a challenge. The expanding literature results in overlapping definitions and approaches, which, although common for nascent concepts, can introduce ambiguities that hinder both theoretical understanding and practical implementations.

Furthermore, these emerging market structures often address specific aspects of the electricity sector in isolation, although they can be interrelated. This raises critical about how these emerging markets integrate into the existing structure of the \gls{EEM} and their interrelation.
These questions motivate our manuscript. We base our research on a combined literature review approach to examine these emerging market concepts' similarities, differences, and interrelations. Our main contribution is a simple organizational classification of established and emerging market structures. We illustrate these emerging market structures fit within the \gls{EEM}, and highlight their underlying commodities.

We structure the rest of this manuscript as follows: In \Autoref{sec:research_appraoch}, we outline our research approach. In \Autoref{sec:eem_structures}, we introduce our simple organizational classification of established and emerging market structures within the \gls{EEM}, emphasizing emerging local markets. In \Autoref{sec:discussion}, we discuss some of the observations (i.e., regulatory ambiguities and market integration, coordination, locational constraints and resource limitations, and trends), while in \Autoref{sec:conclusion} we conclude the manuscript.

\section{Research approach}
\label{sec:research_appraoch}
We use a combined literature review approach to develop a graphical representation of how local markets fit within existing 
\gls{EEM} structures. Our approach integrates principles from two review types while considering established literature review guidelines.  

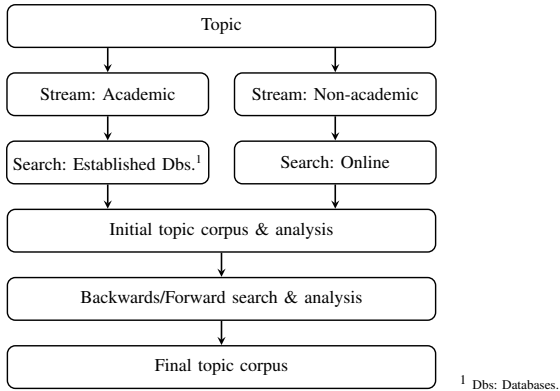
\begin{figure}[h!]
    \centering
    \resizebox{.699\columnwidth}{!}{
     \tikzstyle{box} = [rectangle, rounded corners, minimum height=0.75cm, text centered, draw=black, fill=white, line width=0.25mm]
\tikzstyle{arrow} = [thick,->,>=stealth]

\begin{tikzpicture}[node distance=.5cm]

\small

\node (topic) [box, minimum width=.85\columnwidth] at (0, 0) {Topic}; 
\node (academic) [box, minimum width=.4\columnwidth] at (-2, -1.2) {Stream: Academic};           
\node (nonacademic) [box, minimum width=.4\columnwidth] at (2, -1.2) {Stream: Non-academic};     
\node (dbs) [box, minimum width=.25\columnwidth] at (-2, -2.4) {Search: Established Dbs.\footnote{Dbs:Databases}};          
\node (online) [box, minimum width=.4\columnwidth] at (2, -2.4) {Search: Online};         
\node (initial) [box, minimum width=.85\columnwidth] at (0, -3.6) {Initial topic corpus \& analysis}; 
\node (research) [box, minimum width=.85\columnwidth] at (0, -4.8) {Backwards/Forward search \& analysis}; 
\node (final) [box, minimum width=.85\columnwidth] at (0, -6) {Final topic corpus}; 

\draw [arrow] (topic.south) ++(-2, 0) -- (academic.north);
\draw [arrow] (topic.south) ++(2, 0) -- (nonacademic.north);
\draw [arrow] (academic.south) -- (dbs.north);
\draw [arrow] (nonacademic.south) -- (online.north);
\draw [arrow] (dbs.south) -- ([xshift=-2cm]initial.north);
\draw [arrow] (online.south) -- ([xshift=2cm]initial.north);
\draw [arrow] (initial.south) -- (research.north);
\draw [arrow] (research.south) -- (final.north);

\end{tikzpicture}
    }
    \tiny\textsuperscript{1} Dbs: Databases.
    \caption{Structured search and analysis protocol for combined literature review.   
    }
    
    \label{fig:search_protocol}
\end{figure}

First, we adopt the integrative review approach outlined by \cite{torraco_integrative}. The integrative review approach suits emerging topics well, allowing us to deconstruct them into foundational elements such as history, policy, key concepts, and their relationships. Second, we rely on a narrative review approach based on \cite{PARE2015183} to provide timely, opportunistic insights into the field. To enhance transparency and mitigate common pitfalls associated with integrative and narrative reviews, we follow the guidelines of \cite{SNYDER2019333} and establish a structured search and analysis protocol. While this protocol defines a clear step process, it is not intended to be a quantitative or systematic review due to the topic's complexity. We illustrate our structured search and analysis protocol in \autoref{fig:search_protocol}.  

We apply our protocol to four topics: energy market structures, energy market classifications, local energy communities, and local flexibility markets. We conduct two parallel search streams: one in academic literature and the other in non-academic sources.
For the academic stream, we explore established databases like IEEE, ACM, and Scopus. Our focus is on English-language journals and conference proceedings. We analyze core components such as the title, abstract, and full manuscript when deemed appropriate by the authors.
For non-academic sources, we analyze official documentation from \gls{EU} institutions, including regulations, directives, guidelines, and frameworks. Additionally, we review consultant reports and research projects funded by the \gls{EU}. We retrieve legal and policy documents from EUR-Lex, consultant reports through Google searches, and research project reports from the CORDIS repository.  

Furthermore, we expand our search by analyzing citations (backward search) and identifying subsequent works that cite relevant literature (forward search). Tools like Elicit helped us during these steps. The resulting corpus provides a final topic corpus, which the author team discussed in three meetings.


\section{European Market Structures}
\label{sec:eem_structures}
\subsection{Bird's Eye View}
\label{subsec:EEM_overview}

\begin{figure*}[ht!]
    \centering
   \includegraphics[width=2\columnwidth]{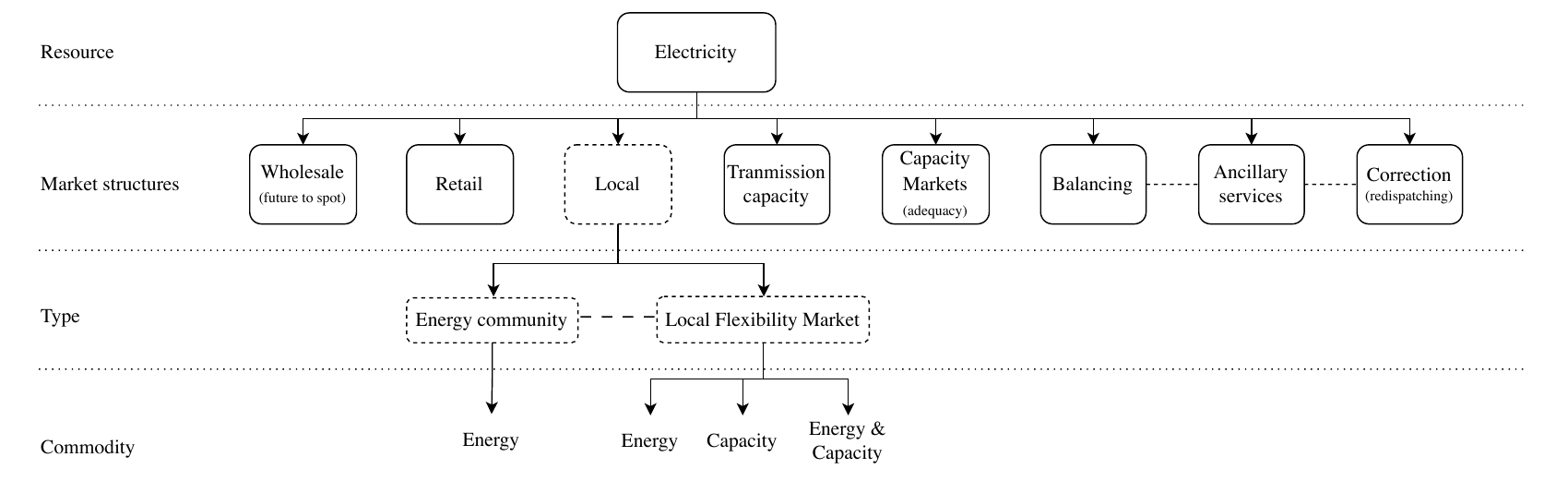}
    \caption{Simple organizational classification of established and emerging market structures.}
    \label{fig:LFM_power_fit}
\end{figure*}

Taking inspiration from the bird's-eye view analogy of~\cite{d2005overall}, we provide a high-level perspective on the current \gls{EEM} structures. We present in \autoref{fig:LFM_power_fit} a simple organizational classification of these markets. 

To understand their fit, we focus on the electricity market structures in the \gls{EU} dealing with electricity as a resource. These have evolved over the past decades to address the challenges of electricity trading and system operation, creating different market structures~\cite{meeus2020evolution}. These market structures cover various functions, from energy procurement, system balancing, ancillary services, and transmission capacity allocation. We split it in our classification to highlight on the left side of the local markets those that target electricity procurement services, while on the right, those more focused on system operation services.

Within electricity procurement services,  wholesale markets operate across various time frames, from forward to spot. Forward and futures markets primarily serve as financial instruments to hedge against short-term price volatility, allowing trading from several years to a month before delivery~\cite{tennet_market_types, emissions_euets_forward_electricity_market}. Then, despite their name, spot markets are not true financial spot markets~\cite{KU_Leuven_fact_sheet}. Instead, they enable shorter-term electricity trading, including day-ahead and intraday horizon, with the latter ones having different modalities (i.e., action or continuous). Retail markets complement these trading markets by facilitating electricity purchases for end consumers through energy suppliers.  

Beyond these electricity procurement services, electricity markets also include structures focused on system operation services. Transmission capacity markets allocate cross-border transmission rights, enabling \glspl{TSO} to manage long-term and short-term network capacity~\cite{entsoe_market_report_2024}. Meanwhile, capacity markets ensure long-term resource adequacy by incentivizing sufficient generation capacity to meet demand~\cite{meeus2020evolution, pototschnig2013capacity}. In parallel, balancing markets provide frequency regulation through primary, secondary, and tertiary reserves, ensuring system stability in real time~\cite{RANCILIO2022111850}. Ancillary services extend beyond frequency control and correction markets like redispatching mechanisms~\cite{meeus2020evolution}. Therefore, we depict connected. Nevertheless, ancillary services can include voltage regulation, black-start capabilities, and interruptible loads, with specific designs varying across countries~\cite{smartnet_d1_1}. However, we chose to illustrate only the main ones.

Local markets emerge at the intersection of electricity procurement and system operation services, functioning as digital platforms that integrate distributed energy resources while addressing local flexibility needs. We categorize these markets into two main types: \glspl{EC} and \glspl{LFM}. \glspl{EC} primarily trade energy as a commodity for remuneration \cite{lopez_european_2024}. In turn, \glspl{LFM} use flexibility services, which, depending on the service, trade energy, capacity, or both as a commodity depending on the service \cite{lfm_spm}.

\subsection{Energy communities}
\label{subsec:energy_communities}

The \gls{EU} defines \glspl{EC} as collective organizations initiated by citizens to engage in various energy-related activities. These activities include energy production, consumption, and storage, managing energy systems, and trading surplus energy in the market~\cite{lopez_european_2024}.  They can operate various energy assets, including solar power systems, energy storage, and heat pumps~\cite{gianaroli_exploring_2024}. They also employ advanced energy management solutions, such as microgrids, energy hubs, and virtual power plants~\cite{lopez_european_2024, haji_bashi_review_2023}. 

Although \glspl{EC} have existed for decades, particularly in remote areas and islands where fuel access is costly and limited, their role has evolved with the rise of decentralized renewable energy production~\cite{lowitzsch_renewable_2020}. The growing adoption of distributed renewables and increasing consumer ownership in energy generation position \glspl{EC} as an emerging model in modern energy market structures~\cite{lowitzsch_renewable_2020}.  

Officially, the \gls{EU} distinguishes between two types of \glspl{EC}: \glspl{CEC} and \glspl{REC}. \glspl{CEC} are legal entities based on voluntary and open participation, controlled by natural persons, local authorities, or small enterprises. Their primary purpose is to provide environmental, economic, or social community benefits rather than financial profits ~\cite{european_parliament_and_of_the_council_directive_2019}. \glspl{REC}, a subset of \glspl{CEC}, operate under an enabling framework designed to facilitate their development~\cite{lowitzsch_renewable_2020}. They focus on renewable energy projects and are controlled by members located near the project sites. While their goals align with those of \glspl{CEC}, they are explicitly tied to renewable energy sources~\cite{european_parliament_and_of_the_council_directive_2018}.  

Despite these definitions, \glspl{EC} share several core features. They require a legal entity as their governing body, participation must be voluntary and inclusive, and their primary objective must extend beyond financial gain. Specific governance rules apply, ensuring that certain participants maintain "effective control"~\cite{frieden_are_2021}. For instance, natural persons, local authorities, and small enterprises can participate in both \glspl{CEC} and \glspl{REC}, whereas large enterprises are restricted from joining \glspl{REC}~\cite{rescoopeu_q_nodate}. These communities often form partnerships between citizens, small businesses, and local authorities, promoting decentralized energy governance~\cite{european_university_institute_robert_schuman_centre_for_advanced_studies_future_2020}.  

Although \gls{EU} directives define \glspl{EC}, \glspl{ME} are responsible for transposing these definitions into national legislation. This has resulted in significant variations in implementation across countries. Some \glspl{ME} do not differentiate between \glspl{CEC} and \glspl{REC}, while others integrate collective self-consumption within the \gls{EC} framework~\cite{frieden_are_2021}.  

Consequently, \glspl{EC} adopt diverse business models \cite{kubli_typology_2023}. Cooperative models are prevalent, emphasizing collective ownership and democratic governance. Additionally, innovative models such as \gls{P2P} energy trading and virtual power plants leverage digital technologies to enhance energy management and market participation. Nevertheless, they always operate from the point of view of energy, being this their commodity, in other words, what is traded (see \autoref{fig:LFM_power_fit}).  

\glspl{EC} function as both market actors and market creators~\cite{trivedi_community-based_2022}. As market actors, they engage with existing electricity markets (see \autoref{fig:LFM_power_fit}), pooling resources to participate in market mechanisms such as selling surplus energy, demand response programs, and grid services. This enables them to access opportunities typically reserved for large market players, contributing to a more decentralized and democratized energy sector~\cite{european_commission_economies_2021}.

Beyond participating in existing markets, \glspl{EC} can act as market creators by establishing new trading platforms, particularly in local contexts~\cite{trivedi_community-based_2022}. Their services range from energy sharing and self-consumption to grid services for stability and optimization~\cite{lopez_european_2024}. For instance, using distributed ledger technologies solutions allows for decentralized energy trading within microgrids, commonly referred to as \gls{P2P} energy trading~\cite{teotia_local_2016}. These \gls{P2P} trading can take three forms: fully decentralized, community-based, and hybrid. In fully decentralized \gls{P2P} models, energy and financial transactions occur directly between members without intermediaries, allowing consumers to set prices and trade energy independently. While this approach maximizes consumer autonomy, it poses scalability challenges and requires significant computational resources. Community-based \gls{P2P} models, by contrast, rely on a central entity that optimizes energy exchange within the community, prioritizing collective efficiency over individual preferences. Hybrid models combine elements of both, enabling cooperation between individual peers and centralized coordination, balancing flexibility with efficiency~\cite{lopez_european_2024}.  
However, \glspl{EC} can also contribute to energy system flexibility. Authors in~\cite{ponnaganti_flexibility_2023} highlight how local \glspl{EC} can provide flexibility services, linking them to energy efficiency measures and \glspl{LFM}. Similarly, authors in~\cite{mendicino_dso_2021} propose a model for integrating \glspl{REC} into flexibility markets, enhancing the coordination of distributed energy resources through nanogrids and small-scale storage.  Meanwhile, authors in~\cite{lopez_european_2024} identify two primary market structures for \gls{EC}-based energy trading: Local Energy Markets and \glspl{LFM}. It is in this context of literature that the interrelation between \glspl{EC} and \glspl{LFM} appears through flexibility services. While with the potential of creating their local markets, these markets might also require flexibility services to provide them internally or beyond the community.

\subsection{Local flexibility markets}
\label{subsec:Flexibility Markets}

\glspl{LFM} have emerged to integrate decentralized flexibility resources into electricity markets while addressing distribution-level congestion \cite{akkouch_congestion_2024}. Their definition varies across contexts, with academic literature viewing them as platforms for trading flexibility within distribution networks \cite{ziras_why_2021}. At the same time, regulatory frameworks emphasize market-based procurement to enhance grid efficiency \cite{european_parliament_and_of_the_council_directive_2019}.  
From a regulatory standpoint, the \gls{EU} \gls{CEP} promotes flexibility procurement through smart meters, digital platforms, and demand-side participation. However, due to their localized nature, \glspl{LFM} pose risks of market power concentration, requiring regulatory safeguards under Directive (EU) 2019/944 \cite{european_parliament_and_of_the_council_directive_2023}.  

A typical \gls{LFM} involves three main participants: flexibility sellers, flexibility buyers, and the \gls{MO} \cite{alavijeh_key_2021, bouloumpasis_local_2022}. Flexibility sellers, often referred to as \glspl{FSP}, include aggregators, generators, consumers, and prosumers. The flexibility buyer, typically a \gls{DSO} but not limited to, assesses grid conditions using congestion forecasts and procures flexibility accordingly. Across various horizons (e.g., long-term to short-term), \glspl{DSO} or \glspl{TSO} may acquire additional flexibility, for instance, for ancillary services or correction ones (see \autoref{fig:LFM_power_fit}). The \gls{MO} ensures bid reception, market clearing, and settlements, facilitating efficient transactions. Some clearing models also incorporate \glspl{BRP} and additional intermediaries \cite{jin_local_2020}.  

\glspl{LFM} primarily mitigate distribution network constraints by utilizing \glspl{DER}, such as \gls{DR}, battery storage, and distributed generation using flexibility services~\cite{mehinovic_quantifying_2025, anaya_role_2021}. We remark that these flexibility services defer grid reinforcements. Flexibility services allow power generation or consumption modifications at specific nodes in predefined time intervals, enhancing local balancing and grid stability \cite{ramos_realizing_2016, wang_mechanism_2023}.  

Flexibility products within \glspl{LFM} vary by activation conditions and timeframes. Real-time flexibility (e.g., "PowerCut Urgent") provides immediate grid support, while baseline-based flexibility ("Drop-by") adjusts consumption relative to expected levels \cite{hennig_congestion_2023}. Flexibility option contracts establish agreements where providers commit to demand reductions under specific conditions, while long-term capacity limitations require consistent electricity use reductions.  

Market structures and coordination models differ across \glspl{LFM}. Some are fully centralized under \glspl{TSO}, while others operate under hybrid \gls{TSO}-\gls{DSO} frameworks or are entirely \gls{DSO}-managed \cite{dronne_local_2021}. Their design depends on national flexibility needs, congestion patterns, and existing market structures.  

Several business models have been proposed to structure \gls{LFM} operations. Authors in \cite{wang_mechanism_2023} describe a framework where \glspl{BRP} and \glspl{DSO} procure flexibility from aggregators, balancing system-wide and local grid requirements. Authors in \cite{heinrich_local_2021} present a market-driven approach with \glspl{DSO}, a market platform, aggregators, and \glspl{DER} owners coordinating congestion management via auctions. Other studies explore strategic frameworks that integrate trading platforms, automated market clearing, and settlement mechanisms to streamline flexibility transactions \cite{jin_local_2020, dronne_local_2021, faia_local_2019}.


\section{Discussion}
\label{sec:discussion}
We highlight key observations and areas for discussion based on our experience drawn from our approach.  

\subsection{Regulatory ambiguities and market integration}

A major challenge for \glspl{EC} and \glspl{LFM} is the lack of clear and consistent definitions across \gls{EU} \glspl{ME}. In the case of \glspl{EC}, the \gls{EU} formally recognizes \glspl{CEC} and \glspl{REC}, yet it does not explicitly define \glspl{EC} as local entities. This ambiguity leads to inconsistent national implementations and legal uncertainty for \glspl{EC} seeking to integrate into existing market structures \cite{energy_communities_repository_barriers_2024}. 
For \glspl{LFM}, regulatory developments are gradually shaping their role as market-based mechanisms for contracting local flexibility \cite{ACER_framework_guidelines, demand_response_network_code_proposal}. They seem to focus on system services at the distribution level \cite{lfm_spm}. While they complement established market structures, such as correction markets (i.e., redispatch), their application extends beyond distribution-level congestion management, as they can also provide flexibility to higher voltage levels. Without precise legal definitions, especially in network codes, \glspl{ME} retain significant discretion in their implementation, creating concept fragmentation across the \gls{EU}. However, their technical implementation heterogeneity is natural as they deal with local problems. Consequently, definitions must also allow for different solutions rather than impose a one-fit-all solution. 

Nevertheless, enhancing their interaction through flexibility services, leveraging complementarities, and addressing shared challenges—while clearly defining their roles within existing markets—could foster the development of more robust and scalable business models.

\subsection{Locational constraints and resource limitations}

While leveraging distributed resources brings social, environmental, and locational benefits, it also introduces constraints. Unlike large energy providers that benefit from economies of scale, \glspl{EC} and \glspl{LFM} operate at a "local" level, where resource availability may be limited. This scarcity, compounded by voluntary participation, may persist over time.  

For \glspl{EC}, sustainability and community-driven fairness are key drivers, yet these priorities make it difficult to compete in price-driven markets \cite{van_den_berghe_community_2022}. Although they can attract consumers in retail markets or secure Power Purchase Agreements (PPAs) through ethical branding, their participation in wholesale markets remains challenging due to cost disadvantages so that business models might disfavor them.  

Similarly, \glspl{LFM} depend on geographically constrained resources, as their current primary function is congestion management at the distribution level \cite{lfm_spm}. This creates variability in design and implementation across countries, reflecting differences in congestion challenges and flexibility needs \cite{dronne_local_2021}. Some \glspl{LFM} focus on long-term flexibility procurement, while others operate in near-real-time \cite{lfm_spm}. Yet, all must aggregate flexibility at a local level, requiring the aggregator to be a facilitator for small resources \cite{akkouch_congestion_2024}. This raises the question:\textit{ how can distributed resources be incentivized to participate in these markets?} Targeted financial mechanisms—such as subsidies for \glspl{EC} engaging in \glspl{LFM} or fixed service-specific contracts in tariff-based solutions—could bridge the gap between regulatory intentions and practical implementation, yet this remains an area for further research.

\subsection{Coordination}

Effective coordination is critical for managing local energy assets and ensuring market efficiency. While synergies exist between \glspl{EC} and \glspl{LFM}, their integration is not mandatory—\glspl{EC} can function independently, and \glspl{LFM} do not inherently rely on them. This raises questions about optimizing their interaction to enhance flexibility and operational efficiency. Furthermore, ensuring interoperability requires standardized frameworks for data exchange and process integration, addressing a broader challenge in smart grid solutions.

Beyond \glspl{EC} and \glspl{LFM}, coordination challenges intensify in multi-market interactions, such as wholesale markets, balancing mechanisms and ancillary services. The broader issue lies in how shared resources and services can be efficiently allocated across different market structures without creating inefficiencies or conflicts. However, these emerging local markets could contribute to addressing this challenge by providing new coordination mechanisms that facilitate resource-sharing across multiple services.

\subsection{Trends: market-based, user-centric, and platforms}

The rise of these local markets is shaped by multiple trends, including market liberalization, user-centric energy systems, and digitalization. Market-driven approaches increasingly dominate, reinforcing the transition toward decentralized, competitive environments where electricity and flexibility are treated as tradable commodities. As regulation continues to emphasize market-based mechanisms, these concepts are likely to evolve into service-oriented solutions. In other words, to enhance their long-term viability, especially given potential resource limitations \glspl{EC} and \glspl{LFM} as platforms will offer various services, where flexibility services will be the main link between these two markets. 

Consumer-centric-driven energy models are also gaining traction. With increasing \gls{DER} adoption, prosumers seek to engage in energy markets beyond self-consumption, facilitated by digital platforms and automation. In \glspl{LFM}, aggregators will likely play an expanding role, pooling resources and introducing new coordination and economic dynamics.  

Digitalization is a key enabler of these trends. Smart metering, automation, and distributed ledger technologies-based trading platforms facilitate dynamic interactions between local energy producers, consumers, and grid operators \cite{rahman_2021, entsoE_plaltforms}. The result is a trend towards creating platforms for \glspl{EC} and \glspl{LFM}.
Thus, further research into market coordination mechanisms, data-sharing frameworks, and cross-border exchanges are potential research streams essential for commercializing these solutions.


\section{Conclusion}
\label{sec:conclusion}
The decentralization of the \gls{EEM} has introduced new market structures, such as \glspl{EC} and \glspl{LFM}, to integrate \glspl{DER} and enhance grid flexibility. However, these markets remain in early development, facing regulatory ambiguities, resource constraints, and coordination challenges.

Our manuscript provides a structured overview of their integration within existing market structures, highlighting their role in bridging electricity procurement and system operation services. \glspl{EC} enable direct community participation in energy markets, while \glspl{LFM} facilitate local flexibility procurement. However, regulatory inconsistencies across \glspl{ME} create fragmentation, limiting market participation and scalability. Their locational constraints and voluntary nature further restrict resource availability, raising concerns about long-term viability.

Their success will depend on regulatory harmonization, business model innovation, and digital infrastructure advancements. Clearer legal definitions and targeted financial incentives could facilitate broader adoption, while platform-based coordination mechanisms will be key to scalability. Future research should focus on adaptive regulations, cross-market coordination, and digital platform integration to enhance the scalability and effectiveness of \glspl{EC} and \glspl{LFM}.

\section*{Acknowledgment}
During the preparation of this work the authors used Grammarly, ChatGPT and Coplitot for proofreading and style improvement. After using these tools/services, the authors reviewed and edited the content as needed and take full responsibility for the content of the publication.


\clearpage

\begin{thebibliography}{10}

\bibitem{fact_sheet_EU}
M.~Ciucci, ``Factsheets - internal energy market,'' 2024.
\newblock Accesed on: 16/01/2025.

\bibitem{8606494}
T.~Gomez, I.~Herrero, P.~Rodilla, R.~Escobar, S.~Lanza, I.~de~la Fuente, M.~L. Llorens, and P.~Junco, ``European union electricity markets: Current practice and future view,'' {\em IEEE Power and Energy Magazine}, vol.~17, no.~1, pp.~20--31, 2019.

\bibitem{collins2002economics}
R.~A. Collins, ``The economics of electricity hedging and a proposed modification for the futures contract for electricity,'' {\em IEEE Transactions on Power Systems}, vol.~17, no.~1, pp.~100--107, 2002.

\bibitem{7521197}
M.~F. Dias, ``Single european energy market: progress and some concerns,'' in {\em 2016 13th International Conference on the European Energy Market (EEM)}, pp.~1--6, 2016.

\bibitem{clean_energy_package}
{European Commission and Directorate-General for Energy}, {\em Clean energy for all Europeans}.
\newblock Publications Office, 2019.
\newblock Accesed on: 16/01/2025.

\bibitem{eu_fit_for_55}
{European Commission}, ``Communication from the commission to the european parliament, the council, the european economic and social committee and the committee of the regions 'fit for 55': delivering the eu's 2030 climate target on the way to climate neutrality.''
\newblock Accessed: 2023-03-04.

\bibitem{Bogdanov2020Full}
D.~Bogdanov, ``Full energy sector transition towards 100

\bibitem{power_systems_SOTA}
M.~{\c C}olak and E.~Irmak, ``A state-of-the-art review on electric power systems and digital transformation,'' {\em Electric Power Components and Systems}, vol.~51, no.~11, pp.~1089--1112, 2023.

\bibitem{cahllenges_power_system}
J.~A. Lopes, A.~Madureira, M.~Matos, R.~Bessa, V.~Monteiro, J.~Afonso, S.~Santos, J.~Catalão, C.~Antunes, and P.~Magalhães, ``The future of power systems: Challenges, trends, and upcoming paradigms,'' {\em Wiley Interdisciplinary Reviews: Energy and Environment}, vol.~9, p.~e368, 12 2019.

\bibitem{Heptonstall2020ASR}
P.~J. Heptonstall and R.~J.~K. Gross, ``A systematic review of the costs and impacts of integrating variable renewables into power grids,'' {\em Nature Energy}, vol.~6, pp.~72--83, 2020.

\bibitem{CABALLEROPENA2022107900}
J.~Caballero-Peña, C.~Cadena-Zarate, A.~Parrado-Duque, and G.~Osma-Pinto, ``Distributed energy resources on distribution networks: A systematic review of modelling, simulation, metrics, and impacts,'' {\em International Journal of Electrical Power \& Energy Systems}, vol.~138, p.~107900, 2022.

\bibitem{GAUR2022113136}
A.~S. Gaur, D.~Z. Fitiwi, M.~Lynch, and G.~Longoria, ``Implications of heating sector electrification on the irish power system in view of the climate action plan,'' {\em Energy Policy}, vol.~168, p.~113136, 2022.

\bibitem{dsm_definition}
P.~Palensky and D.~Dietrich, ``Demand side management: Demand response, intelligent energy systems, and smart loads,'' {\em IEEE Transactions on Industrial Informatics}, vol.~7, no.~3, pp.~381--388, 2011.

\bibitem{eu_regulation_943}
{THE EUROPEAN PARLIAMENT AND THE COUNCIL OF THE EUROPEAN UNION}, ``Regulation (eu) 2019/943 of the european parliament and of the council of 5 june 2019 on the internal market for electricity,'' {\em Official Journal of the European Union}, vol.~L 158, pp.~54--124, 2019-06-14.

\bibitem{european_parliament_and_of_the_council_directive_2019}
{European Parliament and of the Council}, ``Directive ({EU}) 2019/944 of the {European} {Parliament} and of the {Council} of 5 {June} 2019 on common rules for the internal market for electricity and amending {Directive} 2012/27/{EU} (recast) ({Text} with {EEA} relevance.),'' June 2019.

\bibitem{european_commission_joint_research_centre_energy_2020}
{European Commission. Joint Research Centre.}, {\em Energy communities: an overview of energy and social innovation.}
\newblock LU: Publications Office, 2020.

\bibitem{lfm_spm}
S.~{Potenciano Menci} and O.~Valarezo, ``Decoding design characteristics of local flexibility markets for congestion management with a multi-layered taxonomy,'' {\em Applied Energy}, vol.~357, p.~122203, 2024.

\bibitem{ACER_framework_guidelines}
{European Union Agency for the Cooperation of Energy Regulators (ACER)}, ``Framework guideline on demand response.''
\newblock Accessed: 2023-03-04.

\bibitem{demand_response_network_code_proposal}
{The European Commission }, ``Eu dso entity and entso-e proposal for a network code on demand response,'' 2024.
\newblock Accessed: 2025-01-17.

\bibitem{torraco_integrative}
R.~J. Torraco, ``Writing integrative literature reviews: Guidelines and examples,'' {\em Human Resource Development Review}, vol.~4, no.~3, pp.~356--367, 2005.

\bibitem{PARE2015183}
G.~Paré, M.-C. Trudel, M.~Jaana, and S.~Kitsiou, ``Synthesizing information systems knowledge: A typology of literature reviews,'' {\em Information \& Management}, vol.~52, no.~2, pp.~183--199, 2015.

\bibitem{SNYDER2019333}
H.~Snyder, ``Literature review as a research methodology: An overview and guidelines,'' {\em Journal of Business Research}, vol.~104, pp.~333--339, 2019.

\bibitem{d2005overall}
W.~D. D’haeseleer, ``The overall energy issue; a bird’s eye view,'' {\em European Review of Energy Markets}, vol.~1, p.~28, 2005.

\bibitem{meeus2020evolution}
L.~Meeus, {\em The evolution of electricity markets in Europe}.
\newblock Edward Elgar Publishing, 2020.

\bibitem{tennet_market_types}
{TenneT}, ``Market types.'' \url{https://www.tennet.eu/market-types}.
\newblock Accessed: 2025-01-31.

\bibitem{emissions_euets_forward_electricity_market}
{Emissions-EUETS.com}, ``Forward electricity market.'' \url{https://emissions-euets.com/internal-electricity-market-glossary/1472-forward-electricity-market}.
\newblock Accessed: 2025-01-31.

\bibitem{KU_Leuven_fact_sheet}
{KU Leuven Energy Institute}, ``Ei fact sheet: The current electricity market design in europe,'' tech. rep., {KU Leuven Energy Institute}, 2015.

\bibitem{entsoe_market_report_2024}
{ENTSO-E}, ``Entso-e market report 2024,'' tech. rep., European Network of Transmission System Operators for Electricity (ENTSO-E), 2024.
\newblock Accessed: 2025-01-31.

\bibitem{pototschnig2013capacity}
A.~Pototschnig and M.~Godfried, ``Capacity mechanisms and the eu internal electricity market: The regulators' view,'' report, Agency for the Cooperation of Energy Regulators (ACER), 2013.
\newblock Accessed: 2025-01-31.

\bibitem{RANCILIO2022111850}
G.~Rancilio, A.~Rossi, D.~Falabretti, A.~Galliani, and M.~Merlo, ``Ancillary services markets in europe: Evolution and regulatory trade-offs,'' {\em Renewable and Sustainable Energy Reviews}, vol.~154, p.~111850, 2022.

\bibitem{smartnet_d1_1}
G.~Migliavacca, D.~Six, {\em et~al.}, ``Ancillary service provision by res and dsm connected at distribution level in the future power system,'' deliverable d1.1, SmartNet Project, 2016.
\newblock Accessed: 2025-01-31.

\bibitem{lopez_european_2024}
I.~López, N.~Goitia-Zabaleta, A.~Milo, J.~Gómez-Cornejo, I.~Aranzabal, H.~Gaztañaga, and E.~Fernandez, ``European energy communities: {Characteristics}, trends, business models and legal framework,'' {\em Renewable and Sustainable Energy Reviews}, vol.~197, p.~114403, June 2024.

\bibitem{gianaroli_exploring_2024}
F.~Gianaroli, M.~Preziosi, M.~Ricci, P.~Sdringola, M.~A. Ancona, and F.~Melino, ``Exploring the academic landscape of energy communities in {Europe}: {A} systematic literature review,'' {\em Journal of Cleaner Production}, vol.~451, p.~141932, Apr. 2024.

\bibitem{haji_bashi_review_2023}
M.~Haji~Bashi, L.~De~Tommasi, A.~Le~Cam, L.~S. Relaño, P.~Lyons, J.~Mundó, I.~Pandelieva-Dimova, H.~Schapp, K.~Loth-Babut, C.~Egger, M.~Camps, B.~Cassidy, G.~Angelov, and C.~E. Stancioff, ``A review and mapping exercise of energy community regulatory challenges in {European} member states based on a survey of collective energy actors,'' {\em Renewable and Sustainable Energy Reviews}, vol.~172, p.~113055, Feb. 2023.

\bibitem{lowitzsch_renewable_2020}
J.~Lowitzsch, C.~E. Hoicka, and F.~J. van Tulder, ``Renewable energy communities under the 2019 {European} {Clean} {Energy} {Package} – {Governance} model for the energy clusters of the future?,'' {\em Renewable and Sustainable Energy Reviews}, vol.~122, p.~109489, Apr. 2020.

\bibitem{european_parliament_and_of_the_council_directive_2018}
{European Parliament and of the Council}, ``Directive - 2018/2001 - {EN} - {EUR}-{Lex},'' 2018.

\bibitem{frieden_are_2021}
D.~Frieden, A.~Tuerk, A.~R. Antunes, V.~Athanasios, A.-G. Chronis, S.~d’Herbemont, M.~Kirac, R.~Marouço, C.~Neumann, E.~Pastor~Catalayud, N.~Primo, and A.~F. Gubina, ``Are {We} on the {Right} {Track}? {Collective} {Self}-{Consumption} and {Energy} {Communities} in the {European} {Union},'' {\em Sustainability}, vol.~13, p.~12494, Jan. 2021.
\newblock Number: 22 Publisher: Multidisciplinary Digital Publishing Institute.

\bibitem{rescoopeu_q_nodate}
{REScoop.eu}, ``Q\&{A}: {What} are ‘citizen’ and ‘renewable’ energy communities?.''

\bibitem{european_university_institute_robert_schuman_centre_for_advanced_studies_future_2020}
{European University Institute. Robert Schuman Centre for Advanced Studies.}, {\em The future of renewable energy communities in the {EU}: an investigation at the time of the clean energy package.}
\newblock LU: Publications Office, 2020.

\bibitem{kubli_typology_2023}
M.~Kubli and S.~Puranik, ``A typology of business models for energy communities: {Current} and emerging design options,'' {\em Renewable and Sustainable Energy Reviews}, vol.~176, p.~113165, Apr. 2023.

\bibitem{trivedi_community-based_2022}
R.~Trivedi, S.~Patra, Y.~Sidqi, B.~Bowler, F.~Zimmermann, G.~Deconinck, A.~Papaemmanouil, and S.~Khadem, ``Community-{Based} {Microgrids}: {Literature} {Review} and {Pathways} to {Decarbonise} the {Local} {Electricity} {Network},'' {\em Energies}, vol.~15, p.~918, Jan. 2022.
\newblock Number: 3 Publisher: Multidisciplinary Digital Publishing Institute.

\bibitem{european_commission_economies_2021}
{European Commission}, ``Economies of {Energy} {Communities},'' tech. rep., European Commission, 2021.

\bibitem{teotia_local_2016}
F.~Teotia and R.~Bhakar, ``Local energy markets: {Concept}, design and operation,'' in {\em 2016 {National} {Power} {Systems} {Conference} ({NPSC})}, pp.~1--6, Dec. 2016.

\bibitem{ponnaganti_flexibility_2023}
P.~Ponnaganti, R.~Sinha, J.~R. Pillai, and B.~Bak-Jensen, ``Flexibility provisions through local energy communities: {A} review,'' {\em Next Energy}, vol.~1, p.~100022, June 2023.

\bibitem{mendicino_dso_2021}
L.~Mendicino, D.~Menniti, A.~Pinnarelli, N.~Sorrentino, P.~Vizza, C.~Alberti, and F.~Dura, ``{DSO} {Flexibility} {Market} {Framework} for {Renewable} {Energy} {Community} of {Nanogrids},'' {\em Energies}, vol.~14, p.~3460, Jan. 2021.
\newblock Number: 12 Publisher: Multidisciplinary Digital Publishing Institute.

\bibitem{akkouch_congestion_2024}
R.~Akkouch, S.~Potenciano~Menci, and I.~Pavic, ``Congestion {Management} in {European} {Electricity} {Systems} {Through} {Aggregators} and {Local} {Flexibility} {Markets}: {A} {Systematic} {Literature} {Review},'' in {\em 2024 20th {International} {Conference} on the {European} {Energy} {Market} ({EEM})}, (Istanbul, Turkiye), pp.~1--7, IEEE, June 2024.

\bibitem{ziras_why_2021}
C.~Ziras, C.~Heinrich, and H.~W. Bindner, ``Why baselines are not suited for local flexibility markets,'' {\em Renewable and Sustainable Energy Reviews}, vol.~135, p.~110357, Jan. 2021.

\bibitem{european_parliament_and_of_the_council_directive_2023}
{European Parliament and of the Council}, ``Directive ({EU}) 2023/2413 of the {European} {Parliament} and of the {Council} of 18 {October} 2023 amending {Directive} ({EU}) 2018/2001, {Regulation} ({EU}) 2018/1999 and {Directive} 98/70/{EC} as regards the promotion of energy from renewable sources, and repealing {Council} {Directive} ({EU}) 2015/652,'' Oct. 2023.

\bibitem{alavijeh_key_2021}
N.~M. Alavijeh, M.~A.~F. Ghazvini, D.~Steen, L.~A. Tuan, and O.~Carlson, ``Key {Drivers} and {Future} {Scenarios} of {Local} {Energy} and {Flexibility} {Markets},'' in {\em 2021 {IEEE} {Madrid} {PowerTech}}, (Madrid, Spain), pp.~1--6, IEEE, June 2021.

\bibitem{bouloumpasis_local_2022}
I.~Bouloumpasis, N.~Mirzaei~Alavijeh, D.~Steen, and A.~T. Le, ``Local flexibility market framework for grid support services to distribution networks,'' {\em Electrical Engineering}, vol.~104, pp.~401--419, Apr. 2022.

\bibitem{jin_local_2020}
X.~Jin, Q.~Wu, and H.~Jia, ``Local flexibility markets: {Literature} review on concepts, models and clearing methods,'' {\em Applied Energy}, vol.~261, p.~114387, Mar. 2020.

\bibitem{mehinovic_quantifying_2025}
A.~Mehinovic, N.~Suljanovic, and M.~Zajc, ``Quantifying the impact of flexibility asset location on services in the distribution grid: {Power} system and local flexibility market co-simulation,'' {\em Electric Power Systems Research}, vol.~238, p.~111037, Jan. 2025.

\bibitem{anaya_role_2021}
K.~L. Anaya and M.~G. Pollitt, ``The {Role} of {Regulators} in {Promoting} the {Procurement} of {Flexibility} {Services} within the {Electricity} {Distribution} {System}: {A} {Survey} of {Seven} {Leading} {Countries},'' {\em Energies}, vol.~14, p.~4073, July 2021.

\bibitem{ramos_realizing_2016}
A.~Ramos, C.~De~Jonghe, V.~Gómez, and R.~Belmans, ``Realizing the smart grid's potential: {Defining} local markets for flexibility,'' {\em Utilities Policy}, vol.~40, pp.~26--35, June 2016.

\bibitem{wang_mechanism_2023}
S.~Wang, Z.~Hu, and J.~Su, ``Mechanism {Design} of {Blockage} {Management} {Based} on {Local} {Flexibility} {Market} under {High} {Proportion} of {Distributed} {Energy},'' {\em Journal of Physics: Conference Series}, vol.~2442, p.~012018, Feb. 2023.

\bibitem{hennig_congestion_2023}
R.~J. Hennig, L.~J. De~Vries, and S.~H. Tindemans, ``Congestion management in electricity distribution networks: {Smart} tariffs, local markets and direct control,'' {\em Utilities Policy}, vol.~85, p.~101660, Dec. 2023.

\bibitem{dronne_local_2021}
T.~Dronne, F.~Roques, and M.~Saguan, ``Local {Flexibility} {Markets} for {Distribution} {Network} {Congestion}-{Management} in {Center}-{Western} {Europe}: {Which} {Design} for {Which} {Needs}?,'' {\em Energies}, vol.~14, p.~4113, July 2021.

\bibitem{heinrich_local_2021}
C.~Heinrich, C.~Ziras, T.~V. Jensen, H.~W. Bindner, and J.~Kazempour, ``A local flexibility market mechanism with capacity limitation services,'' {\em Energy Policy}, vol.~156, p.~112335, Sept. 2021.

\bibitem{faia_local_2019}
R.~Faia, T.~Pinto, Z.~Vale, and J.~M. Corchado, ``A {Local} {Electricity} {Market} {Model} for {DSO} {Flexibility} {Trading},'' in {\em 2019 16th {International} {Conference} on the {European} {Energy} {Market} ({EEM})}, (Ljubljana, Slovenia), pp.~1--5, IEEE, Sept. 2019.

\bibitem{energy_communities_repository_barriers_2024}
{Energy Communities Repository}, ``{BARRIERS} {AND} {ACTION} {DRIVERS} {FOR} {THE} {DEVELOPMENT} {OF} {DIFFERENT} {ACTIVITIES} {BY} {RENEWABLE} {AND} {CITIZEN} {ENERGY} {COMMUNITIES},'' tech. rep., Energy Communities Repository, 2024.

\bibitem{van_den_berghe_community_2022}
L.~H. G.~J. Van~den Berghe and A.~J. Wieczorek, ``Community participation in electricity markets: {The} impact of market organisation,'' {\em Environmental Innovation and Societal Transitions}, vol.~45, pp.~302--317, Dec. 2022.

\bibitem{rahman_2021}
N.~Rahman, R.~Rabetino, A.~Rajala, and J.~Partanen, ``Ushering in a new dawn: Demand-side local flexibility platform governance and design in the finnish energy markets,'' {\em Energies}, vol.~14, no.~15, 2021.

\bibitem{entsoE_plaltforms}
V.~Charbonnier, C.~Dikaiakos, M.~Foresti, N.~Appleman, M.~Cooper, A.~Hussein, D.~Roberts, and A.~Siriyatorn, ``Review of flexibility platforms,'' tech. rep., ENTSO-E and Frontier Economics Ltd, 2021.

\end{thebibliography}
\end{document}